\documentclass[prd,showpacs,twocolumn,preprintnumbers,amsmath,amssymb]{revtex4}
\usepackage{epsfig}

\def\bea{\arraycolsep .1em \begin{eqnarray}}
\def\eea{\end{eqnarray}}
\def\Tr{{\rm Tr}}

\def\eq#1{(\ref{#1})}

\def\s0#1#2{\mbox{\small{$ \frac{#1}{#2} $}}}
\def\0#1#2{\frac{#1}{#2}}

\def\grgl{\:\hbox to -0.2pt{\lower2.5pt\hbox{$\sim$}\hss}{\raise3pt\hbox{$>$}}\:}
\def\klgl{\:\hbox to -0.2pt{\lower2.5pt\hbox{$\sim$}\hss}{\raise3pt\hbox{$<$}}\:}

\begin{document}

\title{Fixed points of quantum gravity}
 
\vspace{1.5 true cm}
 
\author{Daniel F. Litim}
\affiliation{Theory Division, CERN, CH-1211 Geneva 23\\
\mbox{and School of Physics and Astronomy, University of Southampton,
Southampton SO17 1BJ, U.K.}\\ (Daniel.Litim@cern.ch)}
\preprint{CERN-TH-2003-299, SHEP 03-32}

\begin{abstract}
{Euclidean quantum gravity is studied with renormalisation group
methods.  Analytical results for a non-trivial ultraviolet fixed point
are found for arbitrary dimensions and gauge fixing parameter in the
Einstein-Hilbert truncation. Implications for quantum gravity in four
dimensions are discussed.}
\end{abstract}

\maketitle

Classical general relativity is acknowledged as the theory of gravitational
interactions for distances sufficiently large compared to the Planck
length. At smaller length scales, quantum effects are expected to become
important. The quantisation of general relativity, however, still poses
problems. It is known since long that quantum gravity is perturbatively
non-renormalisable, meaning that an {infinite} number of parameters have to be
fixed to renormalise standard perturbation theory.  It has been suggested that
Einstein gravity may be non-perturbatively renormalisable, a scenario known as
asymptotic safety \cite{Weinberg}. This requires the existence of a
non-trivial ultraviolet fixed point with at most a finite number of unstable
directions. Then it would suffice to adjust a {finite} number of parameters,
ideally taken from experiment, to make the theory asymptotically safe.
Non-perturbative renormalisability has already been established for a number
of field theories \cite{Parisi:1975im}.

The search for fixed points in quantum field theory calls for a
renormalisation group study.  A particularly useful approach is given by the
Exact Renormalisation Group, based on the integrating-out of infinitesimal
momentum shells within a path integral representation of the theory
\cite{ERG-Reviews}.  The strength of the method is its flexibility when it
comes to approximations.  Furthermore, general optimisation procedures are
available \cite{Litim:2000ci}, increasing the domain of validity and the
convergence of the flow. Concequently, the reliability of results based on
optimised flows is enhanced \cite{Litim:2002cf}.

For Euclidean quantum gravity the Exact Renormalisation Group has been made
accessible by Reuter \cite{Reuter:1996cp}. Explicit flow equations have been
constructed using background field techniques
\cite{Reuter:1996cp,Lauscher:2001ya,Lauscher:2002mb}. Diffeomorphism
invariance under local coordinate transformations is controlled by modified
Ward identities, similar to those known for non-Abelian gauge theories
\cite{Freire:2000bq}.  In general, methods originally developed for gauge
theories \cite{Litim:1998qi}, with minor modifications, can now be applied to
quantum gravity.  

So far most studies have been concerned with the Einstein-Hilbert truncation
based on the operators $\sqrt{g}$ and $\sqrt{g}\,R(g)$ in the effective action,
where $g$ is the determinant of the metric tensor $g_{\mu\nu}$ and $R(g)$ the
Ricci scalar. In four dimensions, the high energy behaviour of quantum gravity
is dominated by a non-trivial ultraviolet fixed point
\cite{Lauscher:2001ya,Souma:1999at,Reuter:2001ag}, which is stable under the
inclusion of $R^2(g)$ interactions \cite{Lauscher:2002mb} or non-interacting
matter fields \cite{Percacci:2002ie}.  Further indications for the existence
of a fixed point are based on dimensionally reduced theories
\cite{Forgacs:2002hz} and on numerical studies within simplicial gravity
\cite{Hamber:1999nu}. For phenomenological applications, see
\cite{Bonanno:2000ep}.

In this Letter, we study fixed points of quantum gravity in the approach put
forward in \cite{Reuter:1996cp}, amended by an adequate optimisation
\cite{Litim:2000ci,Litim:2002cf}. The main new result is the existence of a
non-trivial ultraviolet fixed point in the Einstein-Hilbert truncation in
dimensions higher than four, a region which previously has not been
accessible. Analytical results for the flow and its fixed points are given for
arbitrary dimension. The optimisation ensures the maximal reliability of the
result in the present truncation, thereby strengthening earlier findings in
four dimensions. Implications of these results are discussed.

The Exact Renormalisation Group is based on a momentum cutoff for the
propagating degrees of freedom and describes the change of the scale-dependent
effective action $\Gamma_k$ under an infinitesimal change of the cutoff scale
$k$. Thereby it interpolates between a microscopic action in the ultraviolet
and the full quantum effective action in the infrared, where the cutoff is
removed. In its modern formulation, the renormalisation group flow of
$\Gamma_k$ with the logarithmic scale parameter $t=\ln k$ is given by
\begin{equation}\label{ERG}
\partial_t\Gamma_k=
\s012\Tr\left(\Gamma_k^{(2)}+R\right)^{-1}\partial_t R\,.
\end{equation}
The trace stands for a sum over indices and a loop integration, and $R$ (not
be confused with the Ricci scalar) is an appropriately defined momentum cutoff
at the momentum scale $q^2\approx k^2$.  For quantum gravity, we consider the
flow \eq{ERG} for $\Gamma_k[g_{\mu\nu}]$ in the Einstein-Hilbert truncation,
retaining the volume element and the Ricci scalar as independent
operators. Apart from a classical gauge fixing, the effective action is given
by
\begin{equation}\label{EHk}
\Gamma_k=
\0{1}{16\pi G_k}\int d^dx \sqrt{g}\left[-R(g)+2\bar\lambda_k\right]\,.
\end{equation}
In \eq{EHk} we have introduced the gravitational coupling constant $G$
and the cosmological constant $\bar \lambda$.  The Ansatz \eq{EHk}
differs from the Einstein-Hilbert action in $d$ Euclidean dimensions
by the fact that the gravitational coupling and the cosmological
constant have turned into scale-dependent functions. This is a
consequence of the momentum cutoff, expressed in \eq{ERG}. We
introduce dimensionless renormalised gravitational and cosmological
constants $g_k$ and $\lambda_k$ as
\begin{equation}\label{glk}
\begin{array}{l}
g_k=k^{d-2}\, G_k\, \equiv k^{d-2}\, Z^{-1}_{N,k}\ \bar G\\[2ex]
\lambda_k=\,k^{-2}\, \bar\lambda_k\ 
\end{array}
\end{equation}
$\bar G$ and $\bar\lambda$ denote the unrenormalised Newtonian
coupling and cosmological constant at some reference scale
$k=\Lambda$, and $Z_{N,k}$ denotes the wave function renormalisation
factor for the Newtonian coupling. Their flows follow from \eq{ERG} by
an appropriate projection onto the operators in \eq{EHk}. A scaling
solution of the flow \eq{ERG} in the truncation \eq{EHk} corresponds
to fixed points for the couplings \eq{glk}. 

Explicit momentum cutoffs have been provided for the fluctuations in the
metric field in Feynman gauge \cite{Reuter:1996cp}, and for its component
fields in a traceless transverse decomposition in a harmonic background field
gauge with gauge fixing parameter $\alpha$ \cite{Lauscher:2001ya}. In either
case the tensor structure of the regulator is fixed, while the scalar part is
left free.  Here, we employ the optimised cutoff $R_{\rm opt}=(k^2-q^2)\,
\theta(k^2-q^2)$ for the scalar part \cite{Litim:2000ci,Litim:2002cf}.  In the
setup \eq{ERG} -- \eq{glk}, we have computed the flow equation for arbitrary
$\alpha$ and arbitrary dimension.  To simplify the notation, a factor
$1/\alpha$ is absorbed into the definition \eq{glk}. Below we present explicit
formul\ae\ only for the limit $\alpha\to\infty$ where the results take their
simplest form. The general case is discussed elsewhere. The $\beta$-functions
are
\begin{equation}\label{beta-inf}
\begin{array}{l}
\displaystyle
\beta_{\lambda}\equiv
\partial_t\lambda=
\0{P_1}{P_2+4(d+2) g}\\[2ex]
\displaystyle
\beta_{ g}\equiv
\partial_t g=
(d-2+\eta)g=
\0{(d-2)\, g\, P_2}{P_2+4(d+2) g}
\end{array}
\end{equation}
with polynomials $P_{1,2}(\lambda, g;d)$
\bea
\nonumber
\label{P1}
P_1&=&
-16 \lambda^3 
+ 4  \lambda^2 (4 - 10 d\,  g - 3 d^2\,  g + d^3\, g) 
\nonumber\\ &&
+4\lambda (10 d \, g +  d^2\,  g -  d^3 \, g-1) 
\nonumber\\ &&
+ d (2 + d) (d - 16\,  g + 8 d\,  g-3) \, g 
\nonumber\\[1ex]
\label{P2}
\nonumber
P_2&=&
8 ( \lambda^2 -   \lambda -d\, g)+ 2 \,.
\eea
A numerical factor $c_d=\Gamma(\s0d2+2)\,(4\pi)^{d/2-1}$ originating
mainly from the momentum trace in \eq{ERG}, is scaled into the
gravitational coupling $g\to g/c_d$, unless indicated otherwise.  The
graviton anomalous dimension is given by
\bea\label{eta-inf}
\eta&=&     
\0{(d-2)(d+2)\, g}{(d-2) g-2(\lambda-\s012)^2}
\,.
\eea
It vanishes for vanishing gravitational coupling, in two and minus two
dimensions, or for diverging $\lambda$. On a non-trivial fixed point the
vanishing of $\beta_g$ implies $\eta_*=2-d$, and reflects the fact that the
gravitational coupling is dimensionless in $2d$.  This behaviour leads to
modifications of the graviton propagator at large momenta,
e.g.~\cite{Lauscher:2001ya}.  The flows \eq{beta-inf} are finite except on the
boundary $g_{\rm bound}(\lambda)\equiv (2 \lambda - 1)^2/(2d-4)$ derived from
$1/\eta=0$. It signals a breakdown of the truncation \eq{EHk}. Some
trajectories terminate at $g\approx g_{\rm bound}$. The boundary is irrelevant
as soon as the couplings enter the domain of canonical scaling: the limit of
large $|\lambda|$ implies that $\beta_\lambda= -2\lambda$ and $\beta_g=(d-2)g$
modulo subleading corrections. This entails $\eta= 0$. Then the flow is
trivially solved by $g_k=g_\Lambda (k/\Lambda)^{d-2}$ and $\lambda_k
=\lambda_\Lambda (k/\Lambda)^{-2}$, the canonical scaling of the couplings as
implied by \eq{glk}. Here, $\Lambda$ denotes the momentum scale where the
canonical scaling regime is reached. In the infrared limit $g/g_{\rm
bound}(\lambda)\sim (k/\Lambda)^{d+2}$ becomes increasingly small for any
dimension.

In the remaining part, we discuss the fixed points of \eq{beta-inf} and their
implications.  A first understanding is achieved in the limit of vanishing
cosmological constant, where the flow for the dimensionless gravitational
coupling is
\begin{equation}\label{betag0}
\beta_g=\0{(1-4dg)(d-2)g}{1+2(2-d)g}\,.
\end{equation}
The flow \eq{betag0} displays two fixed points, the Gaussian one at $g=0$ and
a non-Gaussian one at $g_*=1/(4d)$. In the vicinity of the non-trivial fixed
point for large $k$, the gravitational coupling behaves as $G_k\approx
g_*/k^{d-2}$.  This behaviour is similar to asymptotic freedom in Yang-Mills
theories.  The anomalous dimension of the graviton \eq{eta-inf} remains finite
and $\le 0$ for all $g$ between the infrared and the ultraviolet fixed point,
because $g_k\le g_{*}<g_{\rm bound}(0)$ for all $d>-2$ and $g_\Lambda\le
g_*$. At the fixed point, universal observables are the eigenvalues of the
stability matrix at criticality, $i.e.$~the critical exponents. The universal
eigenvalue $\theta$ is given by $\partial\beta_g/\partial g|_*=-\theta$.  We
find
\begin{equation}
\theta_{\rm G}=2-d\,, \quad \label{theta0-NG}
\theta_{\rm NG}=2d\ \0{d-2}{d+2} 
\end{equation}
for the Gaussian (G) and the non-Gaussian (NG) fixed point,
respectively.  The eigenvalues at criticality have opposite signs, the
Gaussian one being infrared attractive and the non-Gaussian one being
ultraviolet attractive. They are degenerate in two dimensions. Away
from the fixed points, the flow \eq{betag0} can be solved analytically
for arbitrary scales $k$. With initial condition $g_\Lambda$ at
$k=\Lambda$, the solution $g_k$ for any $k$ is
\begin{equation}\label{g0-general}
\left(\0{g_k}{g_\Lambda}\right)^{-{1}/{\theta_{\rm G}}}
\left(\0{g_*-g_k}{g_*-g_\Lambda}\right)^{-{1}/{\theta_{\rm NG}}}
=
\0k\Lambda\,.
\end{equation}
and $g_*$, $\theta_{\rm G}$ and $\theta_{\rm NG}$ as given
above. Fig.~1 shows the crossover from the infrared to the ultraviolet
fixed point in the analytic solution \eq{g0-general} in four
dimensions (and with $c_4$ in $g$ reinserted).  The corresponding
crossover momentum scale is associated with the Planck mass, $M_{\rm
Pl}=(G)^{-1/2}$. More generally, \eq {g0-general} is a solution at
$\lambda=0$ for arbitrary gauge fixing parameter and regulator. In
these cases, the exponents $\theta$ and the fixed point $g_*$ turn
into functions of the latter. It is reassuring that the eigenvalues
only display a mild dependence on these parameters.

\begin{figure}
\begin{center}
\unitlength0.001\hsize
\begin{picture}(850,800)
\put(380,250){ $ \ln(k/M_{\rm Pl})$}
\put(550,450){ $\eta$}
\put(550,750){ $g$}
\put(120,530){ IR}
\put(730,530){ UV}
\hskip.04\hsize
\epsfig{file=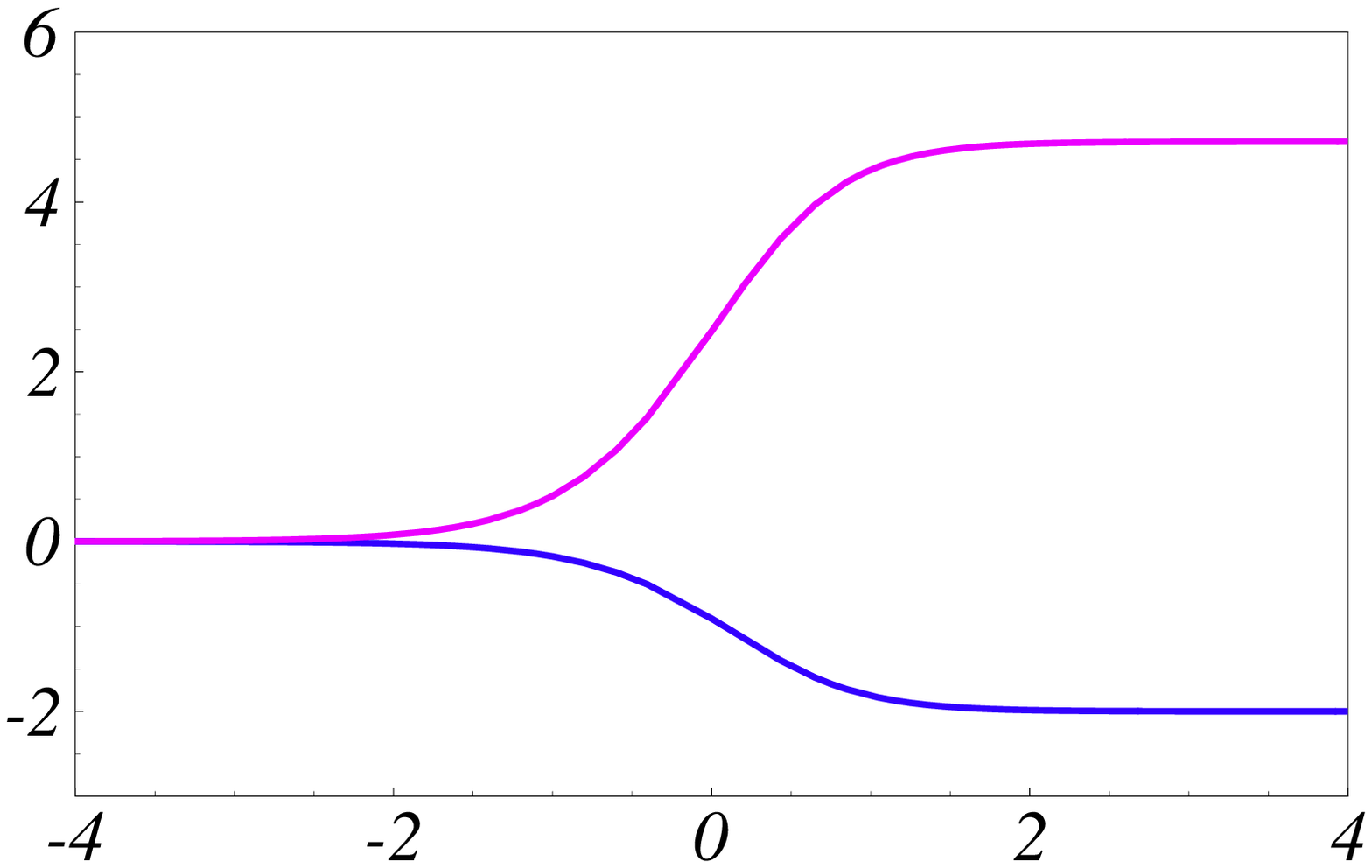,width=800\unitlength}
\end{picture}
\vskip-2cm 
\begin{minipage}{\hsize}
{{\bf Fig.~1:} Running of couplings according to \eq{betag0}, \eq{g0-general}.}
\end{minipage}\vskip-.65cm
\end{center}
\end{figure}

Now we proceed to the non-trivial fixed points implied by the
simultaneous vanishing of $\beta_{ g}$ and $\beta_{ \lambda}$, given
in \eq{glk}.  A first consequence of $P_2=0$ is
$g_*=(\lambda_*-\s012)^2/d$, which, when inserted into $P_1/g_*=0$,
reduces the fixed point condition to a quadratic equation with two
real solutions $(\lambda_*, g_*)\neq 0$ as long as $d\ge d_+$, where
$d_\pm=\s012\left(1\pm\sqrt{17}\right)$\,.  The two branches of fixed
points are characterised by $\lambda_*$ being larger or smaller
$\s012$. The branch with $\lambda_*\ge \s012$ displays an unphysical
singularity at four dimensions and is therefore discarded.  Hence,
\begin{equation}\label{gtilde-inf}
\begin{array}{l}
\displaystyle
\lambda_*=
\0{d^2-d-4-\sqrt{2d(d^2-d-4)}}{2(d-4)(d+1)}\,
\\[2ex]
\displaystyle
 g_*=
\0{2\Gamma(\s0d2+2)(4\pi)^{d/2-1}}{d^2-d-4}\,\lambda_*^2\,.
\end{array}
\end{equation}
In \eq{gtilde-inf}, we have reinserted the numerical factor $c_d$.
The solutions \eq{gtilde-inf} are continuous and well-defined for all
$d\ge d_+\approx 2.56$ and become complex for lower dimensions.  The
critical exponents associated to \eq{gtilde-inf} are derived from the
stability matrix at criticality.  In the most interesting case $d=4$,
the two eigenvalues are a complex conjugate pair
$\theta_\pm\equiv\theta'\pm i\theta''=(5\pm i \sqrt{167})/3$, or
\begin{equation}\label{d=4critexp}
\theta'=1.667\,,\quad\quad\theta''=4.308\,.
\end{equation}
More generally, the eigenvalues remain complex for all dimensions
$2.56<d<21.4$.  In $d=4$, and for general gauge fixing parameter, the
eigenvalues vary between $\theta'\approx 1.5 - 2$ and $\theta''\approx
2.5 - 4.3$.  The range of variation serves as an indicator for the
self-consistency of \eq{EHk}. The result \eq{d=4critexp} and the
variation with $\alpha$ agree well with earlier findings based on
other regulators \cite{Lauscher:2001ya,Souma:1999at}. In the
approximation \eq{betag0}, which does not admit complex eigenvalues,
the ultraviolet eigenvalue reads $\theta_{\rm NG}=8/3$ in $4d$. Both
$\theta_{\rm NG}$ and $\theta'$ agree reasonably well with the
critical eigenvalue $\theta\approx 3$ detected within a numerical
study of $4d$ simplicial gravity with fixed cosmological constant
\cite{Hamber:1999nu}.  In view of the conceptual differences between
the numerical analysis of \cite{Hamber:1999nu} and the present
approach, the precise relationship between these findings requires
further clarification. Still, the qualitative agreement is very
encouraging.

\begin{figure}[t]
\begin{center}
\unitlength0.001\hsize
\begin{picture}(850,800)
\put(-10,770){ \large$g$}
\put(420,250){ \large$\lambda$}
\put(230,430){  {IR}}
\put(195,390){ $(k\to 0)$}
\put(240,720){ crossover region}
\put(630,740){ $g_{\rm bound}$}
\put(270,670){ $(k\sim M_{\rm Pl})$}
\put(610,530){ {UV}}
\put(565,490){ $(k\to \infty)$}
\hskip.04\hsize
\epsfig{file=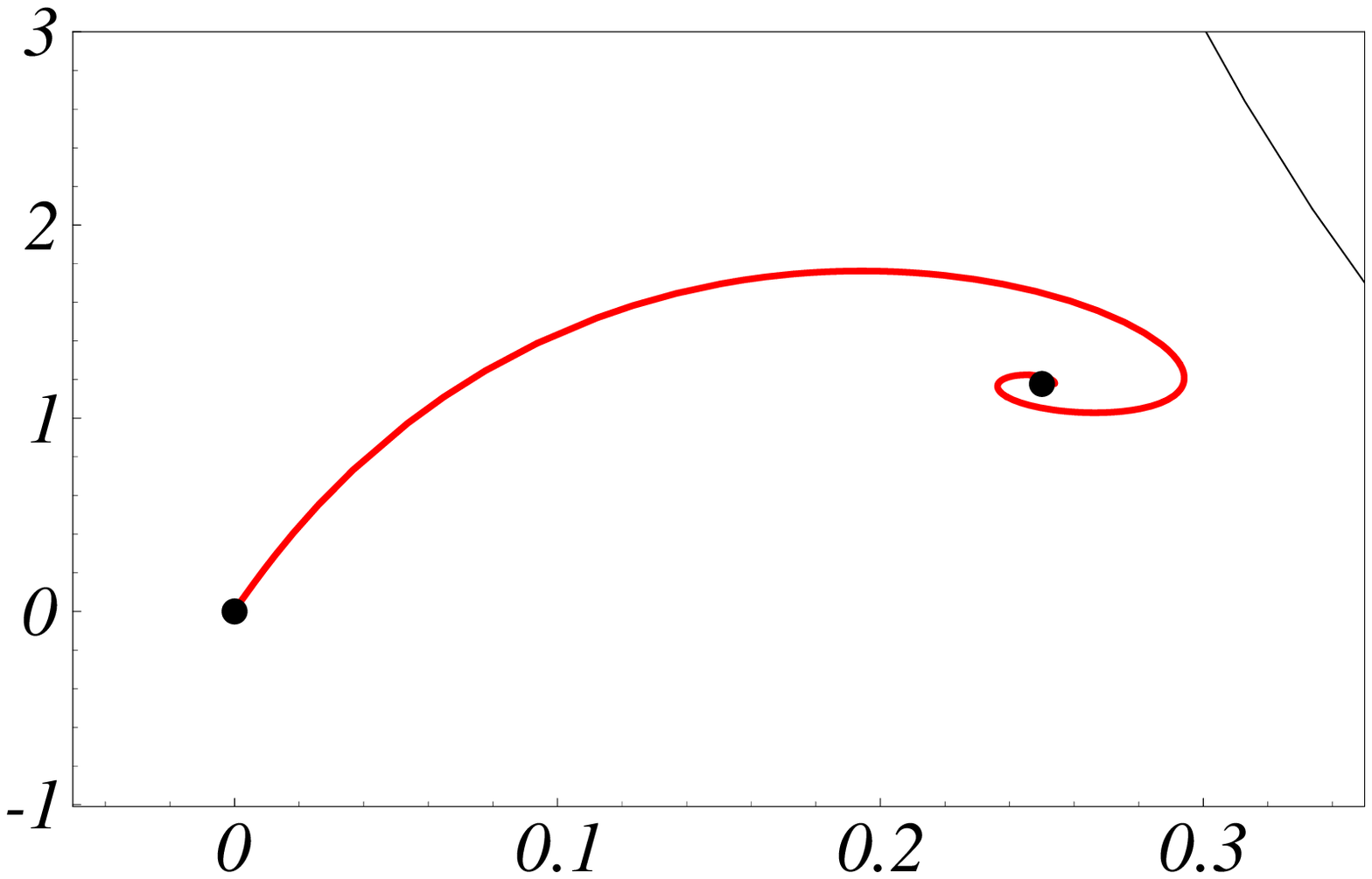,width=800\unitlength}
\end{picture}
\vskip-2.cm 
\begin{minipage}{\hsize}
{{\bf Fig.~2:} The separatrix in four dimensions.}
\end{minipage}\vskip-.6cm
\end{center}
\end{figure}

Next, we discuss the main characteristics of the phase portrait defined
through \eq{beta-inf}.  Finiteness of the flow implies that the line
$1/\eta=0$ cannot be crossed.  Slowness of the flow implies that the line
$\eta=0$ can neither be crossed: in its vicinity, the running of $g$ is
$\beta_g=(d-2)g+{\cal O}(g^2)$, and the gravitational coupling approaches
$g=0$ without ever reaching (nor crossing) it for any scale $k$. Thus,
disconnected regions of renormalisation group trajectories are characterised
by whether $ g$ is larger or smaller $ g_{\rm bound}$ and by the sign of
$g$. Since $\eta$ changes sign only across the lines $\eta=0$ or $1/\eta=0$,
we conclude that the graviton anomalous dimension has the same sign along any
trajectory. In the physical domain which includes the ultraviolet and the
infrared fixed point, the gravitational coupling is positive and the anomalous
dimension negative. In turn, the cosmological constant may change sign on
trajectories emmenating from the ultraviolet fixed point.  Some trajectories
terminate at the boundary $g_{\rm bound}(\lambda)$, linked to the present
truncation (cf.~Fig.~2). The two fixed points are connected by a separatrix.
In Fig.~2, it has been given explicitly in four dimensions (with the factor
$c_4$ in $g$ reinstalled).  Integrating the flow starting in the vicinity of
the ultraviolet fixed point and fine-tuning the initial condition leads to the
trajectory which runs into the Gaussian fixed point.  The rotation of the
separatrix about the ultraviolet fixed point reflects the complex nature of
the eigenvalues \eq{d=4critexp}. At $k\approx M_{\rm Pl}$, the flow displays a
crossover from ultraviolet dominated running to infrared dominated
running. For the running couplings and $\eta$ this behaviour is displayed in
Fig.~3.  The non-vanishing cosmological constant modifies the flow mainly in
the crossover region rather than in the ultraviolet.  A similar behaviour is
expected for operators beyond the truncation \eq{EHk}. This is supported by
the stability of the fixed point under $R^2(g)$ corrections
\cite{Lauscher:2002mb}, and by the weak dependence on the gauge fixing
parameter.  In the infrared limit, the separatrix leads to a vanishing
cosmological constant $\bar\lambda_k=\lambda_k k^2\to 0$. Therefore, it is
interpreted as a phase transition boundary between cosmologies with positive
or negative cosmological constant at large distances. This picture agrees very
well with numerical results for a sharp cut-off flow \cite{Reuter:2001ag},
except for the location of the line $1/\eta=0$ which is non-universal.

\begin{figure}
\begin{center}
\unitlength0.001\hsize
\begin{picture}(850,800)
\put(380,250){ $ \ln(k/M_{\rm Pl})$}
\put(550,540){ $\eta$}
\put(550,680){ $\lambda$}
\put(550,750){ $g$}
\put(120,580){ IR}
\put(730,580){ UV}
\hskip.04\hsize
\epsfig{file=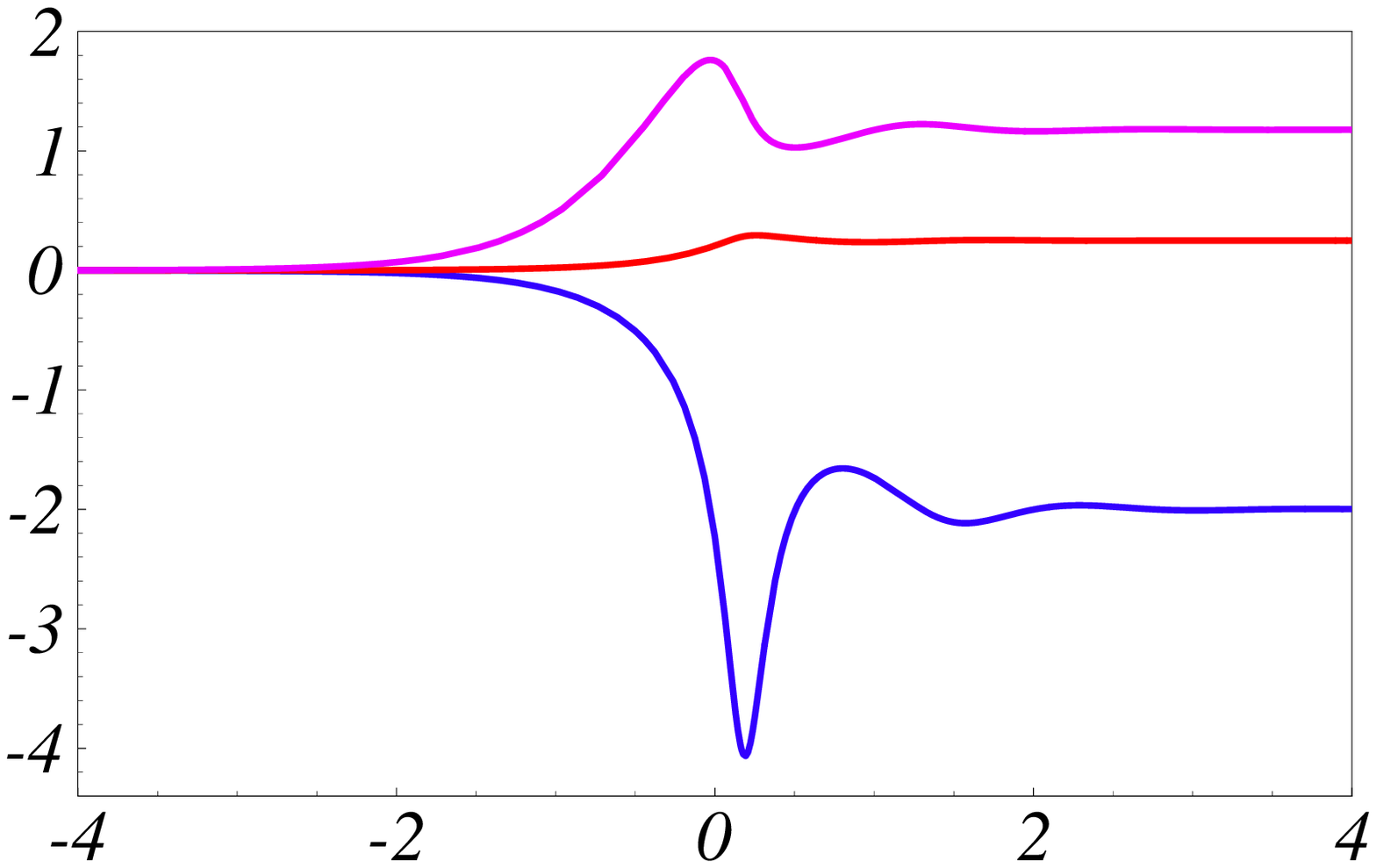,width=800\unitlength}
\end{picture}
\vskip-2.cm 
\begin{minipage}{\hsize}
{{\bf Fig.~3:} Running of couplings along the separatrix.}
\end{minipage}\vskip-.65cm
\end{center}
\end{figure}

Finally, we note that the qualitative picture detailed above persists in
dimensions higher than four. Therefore quantum gravity in higher dimensions
may well be formulated as a fundamental theory.  This consideration is also of
interest for recent phenomenological scenarios based on gravity in extra
dimensions.  In higher dimensions, higher dimensional operators beyond the
truncation \eq{EHk} are likely to be more relevant than in lower
ones. Consequently, the projection of the full flow onto the Einstein-Hilbert
truncation \eq{EHk} and the respective domain of validity are more sensitive
to the cutoff.  A first analysis for specific cutoffs in $d>4$ has revealed
that the fixed point exists up to some finite dimension, where $\lambda_*$
reached the boundary of the domain of validity \cite{Reuter:2001ag}. No
definite conclusion could be drawn for larger dimensions. Based on an
optimised flow, the main new result here is that a non-trivial ultraviolet
fixed point exists within the domain of validity of \eq{beta-inf} for
arbitrary dimension. Furthermore, the fixed point is smoothly connected to its
$4d$ counterpart, and shows only a weak dependence on the gauge fixing
parameter. In the limit of arbitrarily large dimensions, this leads to
$\lambda_*\to 1/2$ and $g_*\to c_d/(2d^2)$ in the explicit solution
\eq{gtilde-inf}. Similar results are obtained for arbitrary gauge fixing
parameter.  Note that $\lambda_*$ approaches its boundary value very slowly,
increasing from $1/4$ to $0.4$ for $d$ ranging from 4 to 40.  Using
\eq{gtilde-inf} and the definition for $g_{\rm bound}$, we derive $g_{\rm
bound}(\lambda_*)/g_*=(2d)/(d-2)$ at the fixed point.  Hence, $g_*$ stays
clear by at least a factor of two from the boundary where $1/\eta$ vanishes,
and the ultraviolet fixed point resides within the domain of validity of
\eq{beta-inf} even in higher dimensions. This stability of the fixed point
also strengthens the result in lower dimensions, including four.

In summary, we have found analytic results for the flow and a non-trivial
ultraviolet attractive fixed point of quantum gravity. Maximal reliability of
the present truncation is guaranteed by the underlying optimisation.  The
fixed point is remarkably stable with only a mild dependence on the gauge
fixing parameter. Furthermore, it extends to dimensions higher than four, a
region which previously has not been accessible.  The qualitative structure of
the phase diagram in the Einstein-Hilbert truncation is equally robust. In
four dimensions the results match with earlier numerical findings based on
different cutoffs.  Hence it is likely that the ultraviolet fixed point exists
in the full theory. We expect that the analytical form of the flow, crucial
for the present analysis, is equally useful in extended truncations. If the
above picture persists in these cases, gravity is non-perturbatively
renormalisable in the sense of Weinberg's asymptotic safety scenario.


\end{document}